\documentclass[12pt,prl,superscriptaddress,showpacs,preprintnumbers,
amsmath,amssymb]{revtex4}

\usepackage{graphicx} 
\usepackage{dcolumn} 

\def\bk{{\bf k}}

\def\bq{{\bf q}}
\def\br{{\bf r}}

\def\b0{{\bf 0}}

\def\bra{\langle}
\def\ket{\rangle}

\def\eps{\epsilon}

\def\Lam{\Lambda}
\def\om{\omega}

\def\sg{\sigma}

\begin{document}

\title{Turning a first order quantum phase transition continuous
 by fluctuations: general flow equations and application to the
 nematic transition in two-dimensional electron systems}

\author{P.~Jakubczyk}
\email{p.jakubczyk@fkf.mpg.de}
\affiliation{Max-Planck-Institute for Solid State Research,
 D-70569 Stuttgart, Germany}
\affiliation{Institute for Theoretical Physics, Warsaw University, 
 Ho\.za 69, 00-681 Warsaw, Poland}
\author{W.~Metzner}
\affiliation{Max-Planck-Institute for Solid State Research,
 D-70569 Stuttgart, Germany}
\author{H.~Yamase}
\affiliation{National Institute for Materials Science, 
 Tsukuba 305-0047, Japan}

\date{\today}

\begin{abstract}
We derive renormalization group equations which allow us to
treat order parameter fluctuations near quantum phase transitions 
in cases where an expansion in powers of the order parameter is 
not possible.
As a prototypical application, we analyze the nematic 
transition driven by a d-wave Pomeranchuk instability in a
two-dimensional electron system. 
We find that order parameter fluctuations suppress the first
order character of the nematic transition obtained at low
temperatures in mean-field theory, so that a continuous
transition leading to quantum criticality can emerge.
\end{abstract}

\maketitle


Numerous itinerant electron systems undergo zero temperature
quantum phase transitions as a function of a tunable control
parameter such as pressure or doping 
\cite{sachdev99,sondhi97,vojta03}.
Most interesting are {\em continuous} transitions associated
with quantum critical fluctuations, which naturally lead to
non-Fermi liquid behavior \cite{loehneysen07}.
However, quantum phase transitions are frequently first order, 
such that critical fluctuations are absent \cite{belitz05}.

In this letter we explore the possibility that a quantum phase
transition which is first order in mean field theory,
turns out to be actually continuous by virtue of order parameter
fluctuations.
As a prototypical example we consider the nematic transition
\cite{kivelson03} driven by a d-wave Pomeranchuk instability 
\cite{yamase00,halboth00} in two dimensions, which is presently
being discussed in the context of cuprates \cite{hinkov04}, 
$\rm Sr_3Ru_2O_7$ \cite{grigera04}, and other correlated 
electron materials.
Mean field studies of tight binding electrons on a square 
lattice indicate that this transition is typically discontinuous 
at zero temperature \cite{kee03,khavkine04,yamase05}.
A continuous transition occurs only at sufficiently high 
temperature, above a tricritical point.
This feature is quite generic since the quartic coupling of 
the Landau free energy describing the transition at zero
temperature is essentially determined by the curvature of the 
electronic density of states at the Fermi level, but with 
opposite sign. 
Due to the logarithmic divergence at the van Hove energy, the 
density of states typically has positive curvature in a broad
window around the van Hove point, leading to a negative quartic
coupling for the order parameter, and hence to a first order
transition.
A continuous transition at zero temperature is found only for
judicious choices of hopping amplitudes and interactions
\cite{yamase05}.

A recent renormalization group (RG) analysis of a quantum
Landau-Ginzburg-Wilson model with $\phi^4$ and $\phi^6$
interactions indicated that order parameter fluctuations can
alter the order of the transition from first (in mean
field theory) to second order \cite{jakubczyk09}.
However, that model is well defined only for a positive 
$\phi^6$-interaction, while, for example, the Landau energy 
for the nematic transition discussed above suffers from negative 
interactions not only at quartic order, but rather at any finite 
order $\phi^{2l}$. The minimum of the Landau energy associated
with the first order transition and the ultimate increase
at large values of $\phi$ cannot be accessed by an expansion
around $\phi = 0$.

In such situations one would like to proceed without resorting
to the usual expansion in powers of the order parameter field. 
The functional RG framework offers the possibility to approximate 
the order parameter interaction by a local potential without 
expanding it in powers of $\phi$ \cite{berges02,delamotte04}.
In the following we will derive the functional flow equations 
for a general quantum Landau-Ginzburg-Wilson model for phase 
transitions in itinerant electron systems, and solve them for 
the specific case of a nematic transition in two dimensions.
We find that fluctuations can indeed transform the first order 
transition obtained in mean field theory into a continuous 
one.


Our starting point is an action for the order parameter 
of the form
\begin{eqnarray}
 {\cal S}[\phi] = 
 \frac{1}{2} \int_q
 \phi_q \left( A_0 \frac{|\omega_{n}|}{|\bq|^{z-2}}
 + Z_0 \bq^{2} \right) \phi_{-q} + {\cal U}[\phi] \; ,
 \label{action}
\end{eqnarray}
where $\phi$ is a scalar order parameter field and $\phi_q$ with
$q = (\bq,\omega_n)$ its momentum representation;
$\omega_n = 2\pi n T$ with integer $n$ denotes the (bosonic) 
Matsubara frequencies.
For the Matsubara sum and momentum integration we use the 
abbreviation
$\int_q = T \sum_{\omega_n} \int \frac{d^dq}{(2\pi)^d}$.
The prefactors $A_0$ and $Z_0$ are positive numbers.
The dynamical exponent $z$ is restricted to values $z \geq 2$.
The potential ${\cal U}[\phi]$ is a functional of $\phi$ which
is local in space and time,
\begin{equation}
 {\cal U}[\phi] = 
 \int_0^{1/T} d\tau \int d^d r \, U(\phi(\br,\tau)) \; ,
\label{potential}
\end{equation}
with a (so far) arbitrary real-valued function $U(\phi)$.
The momenta and frequencies contributing to ${\cal S}[\phi]$
are restricted by an ultraviolet cutoff $\Lambda_0$ to the
region $A_0 \frac{|\omega_{n}|}{|\bq|^{z-2}} + Z_0 \bq^{2} 
\leq \Lambda_0^2$.

An action of the form Eq.~(\ref{action}) can be derived
from a fermionic model by introducing the order parameter as a 
Hubbard-Stratonovich field and subsequently integrating out the
fermions \cite{hertz76,millis93}.
Usually $U(\phi)$ is truncated at quartic order, such that only
a (quadratic) mass term and the local $\phi^4$ interaction is 
kept, which is justified by power counting close to a continuous 
phase transition.
Our essential generalization is that we do not truncate $U(\phi)$
at all, to be able to deal with cases where an expansion in
powers of $\phi$ is not possible. 

We integrate out fluctuations successively by computing the flow
of the effective action $\Gamma^{\Lambda}[\phi]$ from a functional 
RG flow equation. 
The effective action $\Gamma^{\Lambda}[\phi]$ is the generating
functional for one-particle irreducible vertex functions in the
presence of an infrared cutoff $\Lambda$.
The cutoff is implemented by adding a regulator $R^{\Lam}$
to the inverse propagator.
$\Gamma^{\Lambda}[\phi]$ interpolates between the bare action 
$\cal S[\phi]$ for large $\Lambda$ and the final effective
action $\Gamma[\phi]$ in the limit $\Lambda \to 0$.
Its evolution is given by the exact functional flow equation
\cite{wetterich93}
\begin{eqnarray}
 \partial_{\Lambda} \Gamma^{\Lambda}\left[\phi\right]=
 \frac{1}{2}\text{tr}\frac{\dot{R}^{\Lambda}}{\Gamma_2^{\Lambda}
 \left[\phi\right] + R^{\Lambda}} \; ,
 \label{eq:flow_eqn}
\end{eqnarray}
where $\dot{R}^{\Lambda} = \partial_{\Lambda}R^{\Lambda}$, and
$\Gamma_2^{\Lambda}\left[\phi\right] = 
 \delta^2\Gamma^{\Lambda}[\phi]/\delta \phi^2$.
In momentum representation, the trace sums over momenta and 
frequencies: $\text{tr} = \int_q \,$.

The exact effective action is a complicated functional of $\phi$.
We resort to an approximation of the flow based on the following 
ansatz:
\begin{eqnarray}
 \Gamma^{\Lambda}[\phi] = 
 \frac{1}{2} \int_q
 \phi_q \left( A \frac{|\omega_{n}|}{|\bq|^{z-2}}
 + Z \bq^{2} \right) \phi_{-q} + {\cal U}^{\Lambda}[\phi] \; ,
 \label{effaction}
\end{eqnarray}
with $\Lambda$-dependent parameters $A$ and $Z$ and a local 
potential ${\cal U}^{\Lam}[\phi]$ of the form Eq.~(\ref{potential})
with a $\Lambda$-dependent function $U(\phi)$.
A classical version of this ansatz, where Matsubara frequencies
and the dynamical term are absent, has been used previously to 
analyze classical phase transitions \cite{berges02,delamotte04}.
Inserting the ansatz (\ref{effaction}) into the exact flow
equation (\ref{eq:flow_eqn}), and evaluating the resulting
equation for uniform fields,
one obtains the flow equation for $U(\phi)$,
\begin{equation}
 \partial_\Lambda U(\phi) = 
 \frac{1}{2} \, \int_q \, \dot{R}^{\Lambda}(q) \, 
 G^{\Lambda}(q;\phi) \; ,
\label{U-flow}
\end{equation}
where $G^{\Lambda}(q;\phi)$ is the regularized propagator in 
presence of a uniform field $\phi$,
\begin{equation}
 G^{\Lambda}(q;\phi) = 
 \left[ \Gamma_2^{\Lambda}(q;\phi) + R^{\Lambda}(q) \right]^{-1} = 
 \left[ A \frac{|\omega_n|}{|\bq|^{z-2}} + 
 Z \bq^2 + U''(\phi) + R^{\Lambda}(q) \right]^{-1} \; .
\label{propagator}
\end{equation}
Here $U''(\phi)$ is the second derivative of $U(\phi)$ with 
respect to $\phi$, such that the flow equation (\ref{U-flow})
is a partial differential equation.

The evolution of the parameters $A$ and $Z$ parametrizing 
the non-local quadratic term in the effective action
is determined by the flow of $\Gamma_2^{\Lambda}(q;\phi)$,
\begin{eqnarray}
 \partial_\Lambda \Gamma_2^{\Lambda}(q;\phi) &=& 
 \left[\partial_{\phi}^3 U(\phi) \right]^2
 \int_p \dot{R}^{\Lambda}(p) \,
 [G^{\Lambda}(p;\phi)]^2 \, G^{\Lambda}(p+q;\phi) \nonumber \\ 
 && - \, \frac{1}{2} \, [\partial_{\phi}^4 U(\phi)] 
 \int_p \dot{R}^{\Lambda}(p) \, [G^{\Lambda}(p;\phi)]^2 \; .
\label{Gamma2-flow}
\end{eqnarray}
The derivatives $\partial_{\phi}^3 U(\phi)$ and  
$\partial_{\phi}^4 U(\phi)$ correspond to local 3-point and 
4-point vertices, respectively.
We extract the $Z$-factor from the two-point vertex as
$Z = \frac{1}{2d} \left. \Delta_{\bq}
 \left[\Gamma_2^{\Lambda}(q;\phi_0) \right] \right|_{q=0} \,$,
where $\phi_0$ is the position of the global minimum of 
$U(\phi)$. The Laplacian $\Delta_{\bq}$ is defined at 
fixed $R^{\Lambda}$, that is, it does not act on the
regulator $R^{\Lambda}(p+q)$ in $G^{\Lambda}(p+q)$ on the
right hand side of Eq.~(\ref{Gamma2-flow}).
The flow of the $A$-factor could also be extracted from 
$\Gamma_2^{\Lambda}$, but it is of minor importance 
\cite{jakubczyk08} and will be discarded here, that is, 
we set $A = A_0$.

As a regulator we choose a Litim-type \cite{litim01} 
cutoff function
\begin{equation}
 R^{\Lambda}(q) = 
 \left[ Z \left( \Lam^2 - \bq^2 \right) - 
 A|\bq|^{2-z}|\om_n| \right]
 \theta\left[ Z \left( \Lam^2 - \bq^2 \right) - 
 A|\bq|^{2-z}|\om_n| \right] \; ,
\label{Litim_fun}
\end{equation}
which has the convenient feature that the regularized
progagator becomes momentum- and frequency-independent
in the integration region, that is,
$G^{\Lambda}(q;\phi) = [Z\Lambda^2 + U''(\phi)]^{-1}$
for $A |\bq|^{2-z}|\om_n| + Z\bq^2 < Z\Lambda^2$.
The momentum integrations on the right hand side of the 
flow equations can be carried out analytically. 
The frequency summation in Eq.~(\ref{U-flow}) is 
performed numerically.

The flow of $Z$ receives significant contributions only near 
a continuous phase transition at finite temperature, where 
non-Gaussian classical fluctuations generate an anomalous 
scaling dimension (given by the logarithmic derivative of 
$Z$ with respect to $\Lambda$) \cite{jakubczyk08}. 
Note that quantum critical points at $T=0$ are Gaussian 
for $d+z \geq 4$ \cite{hertz76}. 
We can therefore neglect the contributions from $\om_n \neq 0$ 
to the flow of $Z$, and obtain
\begin{equation}
 \frac{d\log Z}{d\log\Lambda} =
 - T [U'''(\phi_0)]^2 \, 
 \frac{S_{d-1}}{(2\pi)^d} \, 
 \frac{\Lambda^{d-6}}{Z^3 (1 + \tilde\delta)^5} \,
 \frac{1}{d} \,
 \left[ 2(1 + \tilde\delta) - \frac{8}{d+2} \right] \; ,
\end{equation}
where $\tilde\delta = U''(\phi_0)/(Z\Lambda^2)$ and
$S_{d-1} = 2\pi^{d/2}/\Gamma(d/2)$ is the area of
the $(d\!-\!1)$-dimensional unit sphere. 


We now apply the flow equations to a specific model system, 
featuring a nematic transition driven by a d-wave Pomeranchuk 
instability.
The model describes tight binding fermions on a square lattice
with an attractive d-wave forward scattering interaction.
Its Hamiltonian is given by \cite{metzner03,dellanna06}
\begin{equation}
 H = \sum_{\bk} \eps_{\bk} n_{\bk} +
 \frac{1}{2L} \sum_{\bk,\bk',\bq} f_{\bk\bk'}(\bq) \,
 n_{\bk}(\bq) \, n_{\bk'}(-\bq) \; ,
\label{f-model}
\end{equation}
where $n_{\bk}(\bq) = \sum_{\sg} 
 c^{\dag}_{\bk-\bq/2,\sg} c^{\phantom\dag}_{\bk+\bq/2,\sg}$
and $L$ is the number of lattice sites.
For nearest and next-to-nearest neighbor hopping the kinetic
energy reads
$\eps_{\bk} = - 2t(\cos k_x + \cos k_y) - 4t' \cos k_x \cos k_y$.
Its van Hove (saddle) points are situated at $\bk = (\pi,0)$ and 
$(0,\pi)$.
The interaction has the form
\begin{equation}
 f_{\bk\bk'}(\bq) = - g(\bq) d_{\bk} d_{\bk'} \; ,
\end{equation}
where $d_{\bk} = \cos k_x - \cos k_y$ is a form factor with
$d_{x^2-y^2}$ symmetry. The coupling function $g(\bq) \geq 0$
has a maximum at $\bq=\b0$ and is restricted to small momentum 
transfers by a momentum cutoff. 
For sufficiently large $g = g(\b0)$ the interaction drives a 
d-wave Pomeranchuk instability leading to a nematic state with 
broken orientation symmetry, which can be described by the 
order parameter
\begin{equation}
 \phi = \frac{g}{L} \sum_{\bk} d_{\bk} \bra n_{\bk} \ket
 \; .
\end{equation}

The mean-field solution of the model (\ref{f-model}) has been
analyzed in a series of articles \cite{kee03,khavkine04,
yamase05}. In the plane spanned by the chemical potential and 
temperature a nematic phase is formed below a dome-shaped 
transition line $T_c(\mu)$ with a maximal transition temperature
near van Hove filling. The phase transition is usually first
order near the edges of the transition line, that is, where
$T_c$ is relatively low, and always second order at the roof
of the dome.

Introducing an order parameter field via a Hubbard-Stratonovich
transformation, integrating out the fermions, and keeping only
the leading momentum and frequency dependences for small $\bq$
and $\om_n/|\bq|$ leads to an action $S[\phi]$ of the form
Eq.~(\ref{action}), with $z=3$ and
\begin{equation}
 U(\phi) = 
 \frac{\phi^2}{2g} - 2T \int \frac{d^2 k}{(2\pi)^2} \,
 \ln\left( 1 + e^{-(\eps_{\bk} - \phi d_{\bk} - \mu)/T}
 \right) \; .
\end{equation}
The coefficients of a Landau expansion of this potential,
$U(\phi) = \frac{a_2}{2} \phi^2 + \frac{a_4}{4!} \phi^4 + \dots$,
are given by $a_2 = g^{-1} - N_2(\mu)$ and
$a_{2l} = - \frac{\partial^{2l-2}}{\partial\mu^{2l-2}} 
N_{2l}(\mu)$ for $l \geq 2$, where $N_{2l}(\mu) = 
 \frac{2}{L} \sum_{\bk} d_{\bk}^{2l} \delta(\eps_{\bk} - \mu)$ 
is the density of states at the Fermi level with a d-wave
form factor \cite{yamase05}. 
Since $N_{2l}(\mu)$ diverges logarithmically at van Hove filling, 
$a_{2l}$ is typically negative for $l \geq 2$.

The coefficients $A_0$ and $Z_0$ in $S[\phi]$ are related to
the d-wave particle-hole bubble and the coupling function 
$g(\bq)$ \cite{dellanna06}.
Note that $A_0$ and $Z_0$ are not involved in the mean-field 
solution of the model Eq.~(\ref{f-model}).
$A_0$ is associated with the Landau damping term in the 
bubble, and depends actually on the orientation of $\bq$.
We will neglect this $\bq$-dependence and insert a constant
$A_0$ corresponding to an angular average.
While the bubble is fully determined by the kinetic energy
$\eps_{\bk}$, the choice of $Z_0$ remains to a large extent
arbitrary since it also depends on the momentum dependence
of $g(\bq)$.
Since momenta $\bq$ in the action $S[\phi]$ correspond to
momentum transfers in the underlying fermionic model, the
ultraviolet cutoff $\Lam_0$ in $S[\phi]$ limits momentum
transfers in $g(\bq)$ to $|\bq| \leq \Lam_0/\sqrt{Z_0}$.

We now present results for the nematic transition as obtained 
from the functional RG.
The following parameters are the same in all quantitative 
plots: $t=1$, $t'=-1/6$, $g=0.8$, $A_0 = 1$, and $Z_0 = 10$.
In Fig.~1 we show an exemplary plot of the evolution of the 
flowing effective potential $U(\phi)$ for $\Lambda$ ranging 
from the ultraviolet cutoff $\Lambda_0 = e^{-1} \approx 0.37$ 
to the final value $\Lambda = 0$. 
\begin{figure}[ht!]
\begin{center}
\includegraphics[width=3.2in]{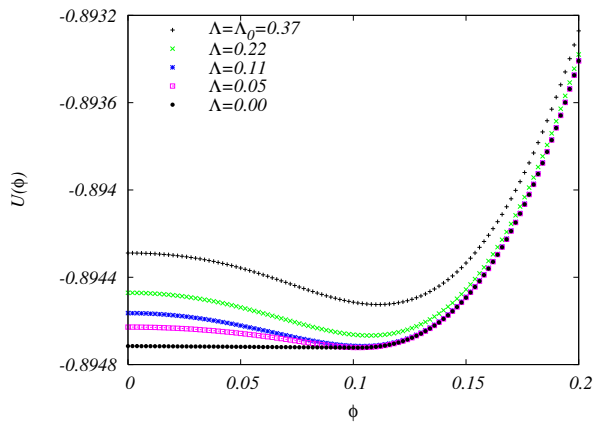}
\caption{(Color online) Effective potential $U(\phi)$ for 
 various values of $\Lambda$ between $\Lambda_0 = e^{-1}$ 
 and $0$, at $\mu = -0.78$ and $T = 0.05$.}
\end{center}
\end{figure}
The initial (mean-field) potential has a minimum at
$\phi_0 = 0.112$.
The final potential exhibits spontaneous symmetry breaking 
with an order parameter $\phi_0 = 0.102$.
Fluctations shift $\phi_0$ toward a slightly smaller value
compared to the mean-field solutions.
Note the flat shape of $U(\phi)$ for $\phi \leq \phi_0$ at
$\Lambda = 0$, which is imposed by the convexity property of 
the grandcanonical potential \cite{goldenfeld92,litim06}.
The final value of $\phi_0$ can be obtained either by 
following the sequence of minima $\phi_0^{\Lam}$ until
$\Lam \to 0$, or by determining the point on the $\phi$-axis
where the flat shape in the final potential $U(\phi)$ 
terminates.

In Fig.~2 we present the transition line between normal and
symmetry-broken phases for two choices of $\Lambda_0$. 
Compared to the corresponding mean-field result, the 
transition temperature is suppressed, with a larger reduction
for larger $\Lambda_0$ (corresponding to a larger phase space
for fluctuations).
\begin{figure}[ht!]
\begin{center}
\includegraphics[width=3.2in]{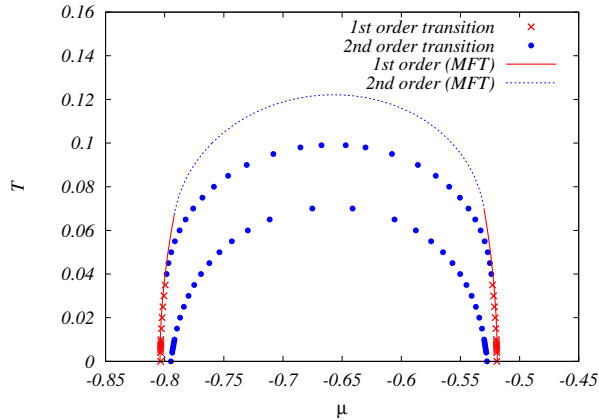}
\caption{(Color online) Critical temperatures versus chemical
 potential for $\Lambda_0 = e^{-1}$ (larger dome with dots and 
 crosses) and $\Lambda_0 = 1$ (smaller dome with dots).
 The mean-field transition line is also shown for comparison.}
\end{center}
\end{figure}
While the phase transition is of first order at low temperatures
in mean-field theory, and also in the presence of fluctuations 
with a cutoff $\Lambda_0 = e^{-1}$,
for $\Lambda_0 = 1$ the transition is continuous down to
$T=0$, leading to quantum critical points $\mu_c$ at the edges 
of the nematic dome.
In Fig.~3 we show the order parameter $\phi_0$ at $T=0$ as a 
function of $\mu$ near the left edge of the transition lines
in Fig.~2 for $\Lam_0 = e^{-1}$ and $\Lam_0 = 1$.
\begin{figure}[ht!]
\begin{center}
\includegraphics[width=3.2in]{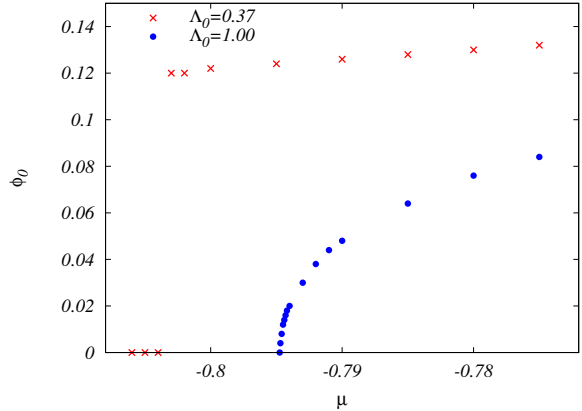}
\caption{(Color online) Order parameter $\phi_0$ as a function
 of $\mu$ at $T=0$ near the left edge of the nematic dome for
 $\Lambda_0 = e^{-1}$ and $\Lambda_0 = 1$.}
\end{center}
\end{figure}
For the case $\Lam_0 = e^{-1}$ the jump of $\phi_0$ indicates
a pronounced first order transition.
By contrast, for $\Lam_0 = 1$ the order parameter rises 
continuously, and the shape of $\phi_0(\mu)$ exhibits the 
expected behavior of a Gaussian quantum critical point.

We note that the mean-field phase diagram is fully determined 
by the parameters $t$, $t'$, and $g$; the size of the phase 
boundary (in units of $t$) depends on $g/t$, but its shape is 
very robust \cite{khavkine04,yamase05}. 
The parameters $A_0$, $Z_0$, and $\Lambda_0$ determine the 
strength and phase space of fluctuations. The suppression of 
$T_c$ and of the first order lines is stronger for larger 
$\Lambda_0$ and smaller $A_0$ and $Z_0$.

In summary, we have presented a renormalization group method
which allows us to treat order parameter fluctuations near
quantum phase transitions in cases where an expansion in
powers of the order parameter $\phi$ is not possible.
We have derived flow equations for the full effective 
potential $U(\phi)$ in a local approximation.
As a first application of the method, we have analyzed
fluctuation effects on the nematic transition driven by a
d-wave Pomeranchuk instability in two-dimensional itinerant
electron systems. 
It turned out that order parameter fluctuations suppress the
first order character of the nematic transition obtained at
low temperatures in mean-field theory, such that a continuous
transition associated with quantum criticality can emerge.
Our flow equations may also be applied to a ferromagnetic 
transition with Ising symmetry and to other Pomeranchuk instabilities.

\vspace*{1cm}

\begin{acknowledgments}
We are grateful to C. Husemann, J. Pawlowski, and P. Strack
for valuable discussions, and to J. Bauer for a 
critical reading of the manuscript.
This work was supported by the German Research Foundation 
through the research group FOR 723.
\end{acknowledgments}


\end{document}